\documentclass[10pt,a4paper]{article} 

\usepackage[utf8]{inputenc}
\usepackage{fontenc}
\usepackage[english]{babel}
\usepackage{csquotes}
\usepackage{amsmath, amsthm, amssymb,mathtools}
\usepackage{graphicx}
\usepackage{mathrsfs}
\usepackage{subcaption}
\usepackage{xargs}
\usepackage{algorithm}
\usepackage[noend]{algpseudocode}
\usepackage{booktabs}
\usepackage{enumitem}
\usepackage{listings}
\usepackage{environ,ifthen,xcolor}
\usepackage{hyperref}
\usepackage{animate}
\mathtoolsset{showonlyrefs}
\definecolor{tuklblue}{RGB}{0,95,140}

\usepackage{geometry}
\usepackage[bordercolor=tuklblue,linecolor=tuklblue,backgroundcolor=tuklblue!20]{todonotes}
\setlist[enumerate,1]{label={\upshape{\roman*)}},itemsep=.25\baselineskip,topsep=.25\baselineskip}
\setlist[enumerate,2]{label={\upshape{\alph*)}},itemsep=.25\baselineskip,topsep=.25\baselineskip}

\DeclareCaptionLabelSeparator{periodspace}{.\ }
\newcommand{\dx}{\mathrm{d}}
\newcommand{\tT}{\mathrm{T}}
\newcommand{\R}{\mathbb{R}}

\newcommandx{\abs}[2][1=\@empty]{#1\lvert #2 #1\rvert}
\newcommandx{\norm}[3][1=\@empty,3=\@empty]{#1\lVert #2 #1\rVert_{#3}}

\DeclareMathOperator*{\argmin}{arg\,min} 

\DeclareMathOperator{\divv}{div}

\DeclareMathOperator{\TV}{TV}

\DeclareMathOperator{\trace}{trace}

\newcommand{\zb}[1]{\mathbf{#1}}

\newboolean{useExternalization}
\setboolean{useExternalization}{false}
\newcommand{\externalFolder}{imgpdf/}
\newcommand{\externalOnly}[1]{
	\ifthenelse{\boolean{useExternalization}}{#1}{}%
}
\NewEnviron{TikZOrPDF}[2][]{
	\ifthenelse{\boolean{useExternalization}}{%
		\tikzsetnextfilename{#2}%
		\BODY
	}{%
	\ifthenelse{\equal{#1}{}}{
		\includegraphics{\externalFolder/#2.pdf}%
	}{
	\includegraphics[width=#1]{\externalFolder#2.pdf}%
}%
}%
}{}%

\externalOnly{
	\usepackage{tikz,pgfkeys,pgfplots,tikz-3dplot}
	\usetikzlibrary{external}
	\tikzexternalize[prefix=\externalFolder]
	\usetikzlibrary{spy,calc,external,arrows}
	\usetikzlibrary{decorations.markings,positioning}
	\tikzstyle{myarrows}=[line width=1mm,draw=gray,-triangle 45,postaction={draw, line width=3mm, shorten >=4mm, -}]
	\usepackage{letltxmacro}
	
	\LetLtxMacro{\oldmissingfigure}{\missingfigure}
	\renewcommand{\missingfigure}[2][]{\tikzexternaldisable\oldmissingfigure[{#1}]{#2}\tikzexternalenable}
	
	\LetLtxMacro{\oldtodo}{\todo}
	\renewcommand{\todo}[2][]{\tikzexternaldisable\oldtodo[#1]{#2}\tikzexternalenable}
	
	\tikzstyle{point}=[inner sep=3ptpt, outer sep=0pt,fill=black]%
}

\title{Examplar-Based Face Colorization Using Image Morphing}

\usepackage{authblk}
\author[1]{Johannes Persch}
\author[1]{Fabien Pierre}
\author[1,2]{Gabriele Steidl}
\affil[1]{Department of Mathematics,
Technische Universit{\"a}t Kaiserslautern 
Paul-Ehrlich-Str.~31, 67663 Kaisers\-lautern, Germany}
\affil[2]{Fraunhofer ITWM, Fraunhofer-Platz 1
67663 Kaiserslautern, Germany}
\date{ May 30, 2017}
\begin{document}

\maketitle
\abstract{Colorization of gray-scale images relies on prior color information.
	Examplar-based methods use a color image as source of such information. 
	Then the colors of the source image are transferred to
	the gray-scale image.
	In the literature, this transfer is mainly guided by texture descriptors. 
	Face images usually contain few texture so that the common approaches frequently fail.
	In this paper we propose a new method based on image morphing. 
	This technique is able to compute a correspondence map between images with similar shapes. 
	It is based on the geometric structure of the images rather than textures which is more reliable for faces.
	Our numerical experiments show that our morphing based approach clearly outperforms state-of-the-art 
        methods.}

\graphicspath{{./Images/}{./}}



\section{Introduction} \label{sec:intro}
Colorization consists in adding color information to a gray-scale images. 
This technique is used for instance by the cinema industry to make old productions more attractive.
As usual we can consider a gray-scale image as luminance channel Y of an RGB image.
The Y channel is defined as a weighted average of the RGB channels: 
$$Y = 0.299R + 0.587G + 0.114B.$$
In addition to the luminance channel, two chrominance channels, called $U$ and $V$,  enable to recover the RGB image. 
Recovering an RGB image from the luminance channel alone is an ill-posed problem 
and requires additional information. 
This information is provided in the literature in two different ways, namely by {\it manual} or {\it exemplar-based} methods.
In the first one the user augments the image by some color strokes as basis for the algorithm to compute the color of each pixel. 
The colorization of a complex scene by a manual prior can be a tedious work for the user. 
In the second approach a color image is used as a source of information. 
Here the results strongly depend on the choice of the image. 
Therefore it is often called {\em semi-automatic}. 

In this paper we focus on the exemplar-based methods. 
A common back-bones of these techniques is the matching of the images. 
First, the {\it source} image is transformed to a gray-scale image 
which is compared with the input gray-scale image called {\em target}. 
The main issue of the exemplar-based colorization consists in matching the pixels of the source and the target gray-scale images. 
The basic hypothesis in the literature is that the color content is similar in similar texture patches.
Then the main challenge is the choice of appropriate texture descriptors. 

In the literature, the exemplar-based methods come along with various matching procedures.
Some of them are done after an automatic segmentation, whereas others use local information. 
The seminal paper on exemplar-based colorization by Welsh et al.~\cite{welsh2002transferring}
was inspired by the texture synthesis of Efros and Leung~\cite{efros1999texture} and
uses basic descriptors for image patches (intensity and standard-deviation) to describe the local texture. 
Pierre et al.~\cite{pierre2015luminance} proposed an exemplar-based framework 
based on various metrics between patches to produce a couple of colorization results. 
Then a variational model with total variation like regularization is applied 
to chose between the different results in one pixel with a spatial regularity assumption. 
The approach of Irony et al.~\cite{irony2005colorization} is built
on the segmentation of the images by a mean-shift algorithm. 
The matching between segments of the images is computed from DCT descriptors which analyse the textures. 
The method of Gupta et al.~\cite{gupta2012image} is rather similar. 
Here, an over-segmentation (SLIC, see, e.g.,~\cite{achanta2012slic}) 
is used instead of the mean-shift algorithm. The comparison between textures in done by SURF and Gabor features. 
Chen et al.~\cite{chen2004grayscale} proposed a segmentation approach 
based on Bayesian image matching 
which can also deal with smooth images including faces. 
The authors pointed out  that the colorization of faces is a particular hard problem. 
However, their approach uses a manual matching between objects to skirt the problem of smooth parts. 
Charpiat et al.~\cite{charpiat2010automatic} ensured spatial coherency without segmenting, but their
method involves many complex steps. 
The texture discrimination is mainly based on SURF descriptors. 
In the method of Chia et al.~\cite{chia2011semantic}, the user has manually  to segment and label the objects
and the algorithm finds similar segments in a set of images available in the internet.
Recently a convolutional neural network (CNN) has been used for colorization by Zhang et al.~\cite{zhang2016colorful} with promising results. 
Here the colorization is computed from a local description of the image. 
However, no regularization is applied to ensure a spatial coherence. This produces ,,halo effects'' near strong contours. 
All the described methods efficiently distinguish textures and possibly correct them with variational approaches, 
but fail when similar textures have to be colorized with different colors. 
This case arises naturally for face images. 
Here the smooth skin is considered nearly as a constant part. 
Thus, when the target image contains constant parts outside the face, the texture-based methods fail. 

In this paper we propose a new technique for the colorization of face images guided by image morphing.
Our framework relies on the hypothesis 
that the global shape of faces is similar. 
The matching of the two images is performed by computing a morphing map 
between the target image and the gray-scale version of the source image. 
The morphing or metamorphosis approach was first proposed by Miller and Younes \cite{MY2001}.
In this paper we build up on a time discrete version suggested by Berkels et al. \cite{BER15}.
In contrast to these authors we apply a finite difference approach which solves
a registration problem in each iteration step.
Having the morphing map available, the chrominance channels can be transported by this map
to the target image, while preserving its luminance channel.
This gives very good results and outperforms state-of-the-art methods.
For some images we can further improve the quality by applying
a variational post-processing step. This was done in two of our numerical examples.
Our variational model incorporates a total variation like regularization term which takes the edges of the target image into account.
This was also proposed by one of the authors of this paper in~\cite{pierre2015luminance}.
However the method is accomplished by adapting a procedure of Deledalle et al.~\cite{deledalle2017clear}
to keep the contrast of the otherwise biased total variation method.

The outline of the paper is as follows:
In Section~\ref{sec:morphing} we sketch the ideas of the morphing approach.
In particular we show how the morphing map is computed with an alternating minimization algorithm. 
Section~\ref{sec:color} deals with the color transfer. 
Having the morphing map at hand the transfer of the chrominance values is described in Subsection~\ref{subsec31}. 
Sometimes it appears to be useful to apply a variational model with a 
modified total variation regularization as post-processing step
to remove possible spatial inconsistencies. This step is described in Subsection~\ref{subsec32}. 
Numerical experiments demonstrate the very good performance of our algorithm in Section~\ref{sec:numerical}. 
The paper ends with conclusions in Section \ref{sec:conclusions}.

\section{Image Morphing}\label{sec:morphing}
Our colorization method is based on image morphing, also known as image metamorphosis
which we briefly explain in this section. For a detailed description we refer to \cite{PS2017}.
The morphing or metamorphosis approach was first proposed by Miller and Younes \cite{MY2001}, see also \cite{Younes2010}
with a comprehensive analysis by Trouv\'e and Younes \cite{TY2005}.
The basic idea consists in considering images as objects lying on a Riemannian manifold
with a Riemannian  metric that takes diffeomorphisms between images into account.
Then the source and the target image are the starting and end points, respectively, of curves lying on the manifold.
The method aims to find the shortest path between the images, i.e., the geodesic joining them
by minimizing the path energy. By definition of the Riemannian metric this provides not only a sequence of images along the path but also
the diffeomorphism which moves the image intensities pixel-wise along the path.
So far we are in a continuous setting, i.e., images are functions on a domain $\Omega \subset \mathbb R^2$ 
mapping into $\mathbb R$ and the curves evolve continuously in time.
Recently a time-discrete path approach was proposed by Berkels et al.~\cite{BER15}.
This is the starting point of our work which we describe next.

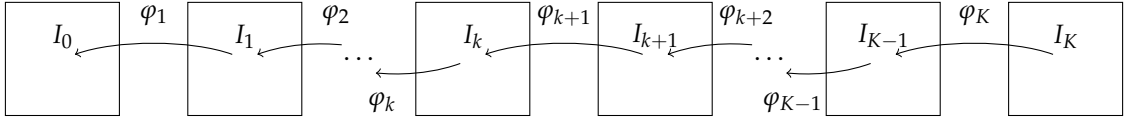
\begin{figure}[t]
	\centering
\begin{TikZOrPDF}{ImageSeq}
	\begin{tikzpicture}[scale=1.5]	
	\draw (0,0) rectangle (1,1);
	\node at (0.5,.7) (I0) {$I_0$};
	\draw[<-,shorten <= 5,shorten >= 5] (.5,.5) ..controls +(20:.5)  and +(160:.5).. (2.1,.5) node [pos = 0.5,label = 90:$\varphi_1$](phi0) {};
	\draw (1.6,0) rectangle (2.6,1);
	\node at (2.1,.7) (I1) {$I_1$};
	\draw[<-,shorten <= 5] (2.1,.5) ..controls +(20:.5)  and +(180:.025).. (2.95,.625) node [pos = 0.8,label = 90:$\varphi_2$](phi1) {};
	
	\node at (3.1,.5) (d1) {$\Huge{\dots}$};
	\draw[<-,shorten >= 5] (3.25,.375) ..controls +(0:.025)  and +(200:.5).. (4.1,.5) node [pos = 0.2,label = -90:$\varphi_{k}$](phikm1) {};
	\draw (3.6,0) rectangle (4.6,1);
	\node at (4.1,.7) (Ik) {$I_k$};
	\draw[<-,shorten <= 5,shorten >= 5] (4.1,.5) ..controls +(20:.5)  and +(160:.5).. (5.7,.5) node [pos = 0.5,label = 90:$\varphi_{k+1}$](phik) {};
	\draw (5.2,0) rectangle (6.2,1);
	\node at (5.7,.7) (Ik1) {$I_{k+1}$};
	\draw[<-,shorten <= 5] (5.7,.5) ..controls +(20:.5)  and +(180:.025).. (6.55,.625) node [pos = 0.8,label = 90:$\varphi_{k+2}$](phik1) {};
	\node at (6.7,.5) (Ik) {$\Huge{\dots}$};
	\draw[<-,shorten >= 5] (6.85,.375) ..controls +(0:.025)  and +(200:.5).. (7.7,.5) node [pos = 0.2,label = -90:$\varphi_{K-1}$](phKkm1) {};
	\draw (7.2,0) rectangle (8.2,1);
	\node at (7.7,.7) (Ik) {$I_{K-1}$};
	\draw[<-,shorten <= 5,shorten >= 5] (7.7,.5) ..controls +(20:.5)  and +(160:.5).. (9.3,.5) node [pos = 0.5,label = 90:$\varphi_{K}$](phiK) {};
	\draw (8.8,0) rectangle (9.8,1);
	\node at (9.3,.7) (Ik1) {$I_{K}$};
	\end{tikzpicture}
\end{TikZOrPDF}
	\caption{Illustration of the image path and the diffeomorphism path. \label{fig:sequence}}
\end{figure}

Let $\Omega \subset \mathbb R^2$ be an open, bounded domain with  Lipschitz continuous boundary.
We are given a gray-value template image $I_{\text{temp}}: \Omega \rightarrow \mathbb R$ 
and a target image
$I_{\text{ tar}}: \Omega \rightarrow \mathbb R$ 
which are supposed to be continuously differentiable and compactly supported. 
For $K \ge 2$ set
\begin{equation}\label{begin_end}
I_0 \coloneqq I_{\text{temp}}, \quad I_K \coloneqq I_{\text{tar}}.
\end{equation}
In our application, the template image will be the luminance channel of the color source image
and the target image the gray-scale image we want to colorize.
We want to find a sequence of $K-1$ images ${\zb I}$ together with a sequence of diffeomorphisms $\boldsymbol{\varphi}$ on $\Omega$,i.e.,
$${\zb I} \coloneqq (I_1,\ldots,I_{K-1}), \quad \boldsymbol{\varphi} \coloneqq (\varphi_1,\ldots ,\varphi_K),$$ 
such that 
$$I_k \approx I_{k-1} \circ \varphi_k \quad \mbox{for all} \; x \in \Omega,$$ 
see Figure \ref{fig:sequence},
and the deformations $\varphi_k$ have a small
linearized elastic potential defined below. 
To this end, we suppose for $k=1,\ldots,K$ that $\varphi_k$ is related to
the displacement $v_k$ by
\begin{equation}\label{disp}
\varphi_k(\zb x) = \zb x - v_k(\zb x) = 
\begin{pmatrix} 
x - v_{k,1} (\zb x)\\
y - v_{k,2} (\zb x)
\end{pmatrix}, 
\quad \zb x = (x,y)^\tT \in \Omega,
\end{equation}
and set $\zb v \coloneqq (v_1,\ldots,v_K)$.
The (Cauchy) strain tensor of the displacement $v = (v_1,v_2)^\tT : \Omega \rightarrow \mathbb R^2$ is defined by
$$
\varepsilon (v) \coloneqq \frac12 (\nabla v + \nabla v^\tT) 
= 
\begin{pmatrix} 
\partial_x v_1 & \frac12 (\partial_y v_1 + \partial_x v_2)\\
\frac12 (\partial_y v_1 + \partial_x v_2) & \partial_y v_2
\end{pmatrix},
$$
where $\nabla v$ denotes the Jacobian of $v$. 
The linearized elastical potential is given by
\begin{equation} \label{potential}
\mathcal{S}(v) \coloneqq \int_{\Omega} \mu \trace \left( \varepsilon^\tT (v) \varepsilon ( v ) \right)
+
\frac{\lambda}{2}\trace \left(\varepsilon(v)\right)^2 \, \dx \zb x ,
\end{equation}
where $\mu,\lambda > 0$.
Then we want to minimize
\begin{align} \label{main}
\mathcal{J}(\zb I,\zb v) &\coloneqq  \sum_{k=1}^K \int_{\Omega} | I_k - I_{k-1} \circ \varphi_k |^2  + \mathcal{S}(v_k) \, \dx \zb x ,\quad
\mbox{subject to} \quad I_0 = I_{\text{ temp}}, \; I_K = I_{\text{tar}}.
\end{align}
This functional was introduced and analyzed by Berkels et al. in \cite{BER15}.
The minimizer $(\zb I, \zb v)$ of the functional provides us with both a sequence of images $\zb I$ 
along the approximate geodesic path and a sequence of displacements $\zb v$ 
managing the transport of the gray values through this image sequence. 
Note that the term
$\mathcal{S}(v)$ may be accomplished by a higher order derivative 
$$
\int_{\Omega} \lvert D^m v_k(x)\rvert^2\dx x , \quad  m >2,
$$
which ensures in the time continuous setting that $\varphi$ is indeed a diffeomorphism,  see \cite{DGM98}.

For finding a minimizer of \eqref{main} we alternate the minimization over $\zb I$ and $\zb v$ as also proposed in \cite{BER15}:
\begin{itemize}
	\item[1.] Fixing $\zb I$ and minimizing over $\zb v$ leads to the following $K$ single registration problems:
	\begin{equation} \label{fixI}
	\argmin_{v_k} \mathcal{J}_{\zb I}(v_k) \coloneqq \int_{\Omega} | I_k - I_{k-1} \circ \varphi_k |^2  + \mathcal{S}(v_k) \, \dx \zb x, \quad k=1, \ldots,K
	\end{equation}
	where $\varphi_k$ is related to $v_k$ by \eqref{disp}. 
	\item[2.] 
	Fixing $\zb v$, resp., $\boldsymbol{\varphi}$ leads to solving the following image sequence problem
	\begin{equation} \label{fix_u}
	\argmin_{{\zb I}} \mathcal{J}_{\boldsymbol{\boldsymbol{\varphi}}}(\zb I) \coloneqq  \sum_{k=1}^K \int_{\Omega} | I_k - I_{k-1} \circ \varphi_k |^2 \, \dx \zb x. 
	\end{equation}
	This can be done via the linear system of equations
	arising from Euler-Lagrange equation of the functional which
	we describe in the Appendix \ref{app}.
\end{itemize}

Dealing with digital images which map from a rectangular image grid $\{1,\dots,n_1\} \times \{1,\dots,n_2\}$
into $\mathbb R$ rather than from a continuous domain $\Omega \subset \mathbb R^2$, 
we need a spatial discretization of the functional \eqref{main}.
The authors of \cite{BER15} propose to use a finite element approach 
for the spatial discretization.
Finite element methods are highly flexible and
can be also applied, e.g.,~for shape metamorphosis. 
However, having the rectangular structure of the image grid in mind, we propose to use finite
differences for the spatial discretization. 
This has also the advantage that we can build up in the registration step 1 of our alternating algorithm
on methods proposed by Haber and Modersitzki  \cite{HM06}, see also \cite{Mod2004}.
As usual in optical flow and image registration we work on a staggered grid and apply a coarse-to-fine strategy.
For the concrete spatial discretization of the differential operators in \eqref{main}, the solution of the registration problem and 
detailed numerical issues concerning the
computation of the morphing map $\varphi$ we refer to  \cite{PS2017}.

\section{Face Colorization}\label{sec:color}

In this section, we describe a method to colorize a gray-scale image based on the morphing map between the luminance channel
of a source image (template) and the present image (target).
The approach is based on the transfer of the chrominance channels from the source image to the target one.
Further, we propose a variational method with a total variation based regularization as a post-processing step to remove possible artifacts. 

\subsection{Color Transfer}\label{subsec31}

The color transfer will be done in the YUV color space.
While in RGB images the color channels are highly correlated, the 
YUV space shows a suitable decorrelation between the luminance channel $Y$ and the two chrominance channels $U,V$.
The transfer from the RGB space to YUV space is given by
\begin{equation}\label{yuvrgb}
\left(\begin{array}{c}
Y \\ 
U \\ 
V
\end{array} \right)
=
\left( \begin{array}{ccc}
0,299 & 0,587 & 0,114 \\ 
-0,14713 & -0,28886 & 0,436 \\
0,615 & -0,51498 & -0,10001
\end{array} \right)
\left(\begin{array}{c}
R \\ 
G \\ 
B
\end{array} \right) .
\end{equation}
Most of the colorization approaches are based on the hypothesis that the target image is the luminance channel of the desired image. 
Thus, the image colorization process is based on the computation of the unknown chrominance channels. 

\paragraph{Luminance Normalization}
The first step of the algorithm consists in transforming the RGB source image to the $YUV$ image by \eqref{yuvrgb}.
The range of the target gray-value image and the $Y$ channel of the source image may differ 
making the meaningful comparison between these images not possible. 

To tackle this issue, most of state-of-the art methods use a technique called \emph{luminance remapping} which 
was introduced in~\cite{hertzmann2001image}. This affine mapping between images 
which aims to fit the average and the standard deviation of the target and the template images  is 
defined as
\begin{equation}
I_{\text{temp}} \coloneqq \sqrt{\dfrac{\operatorname{var}(I_{\text{tar}})}{\operatorname{var}(I_{Y})}}\left( I_{Y} - \operatorname{mean}(I_{Y}) \right) + \operatorname{mean}(I_{\text{tar}}), 
\end{equation}
where $\operatorname{mean}$ is the average of the pixel values, and $\operatorname{var}$ is the empirical variance. 

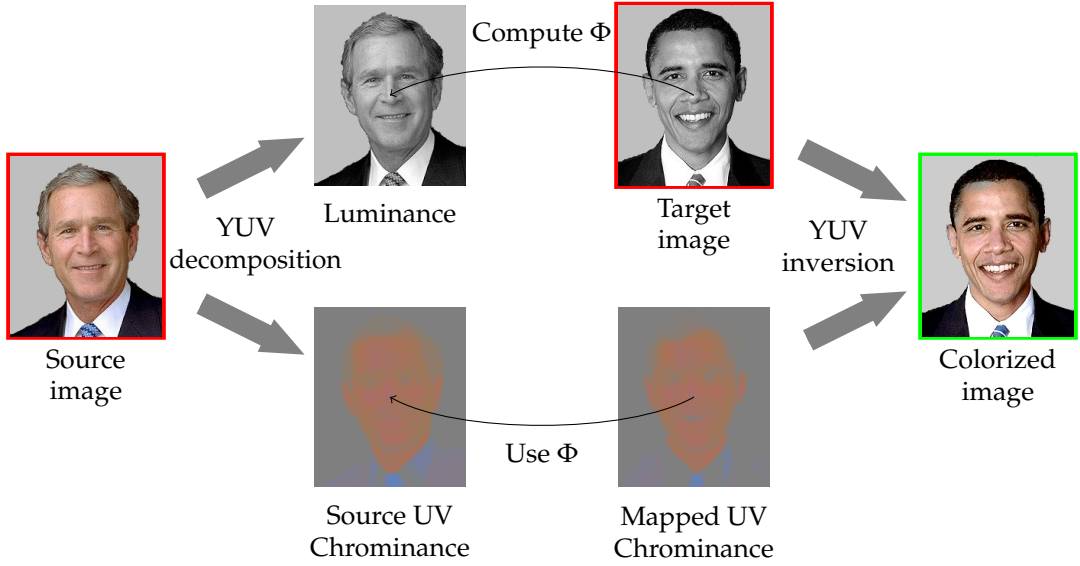
\begin{figure}[t]
	\centering
	\begin{TikZOrPDF}{scheme}
	\begin{tikzpicture}[scale=1]	
	\node[draw,line width=1mm,color = red,inner sep = 0] at (0,0) (S) {\includegraphics[width = 2cm]{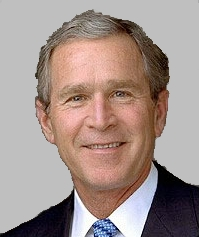}};
	\node[draw,line width=1mm,color = green,inner sep = 0,outer sep = 2mm] at (12,0) (R) {\includegraphics[width = 2cm]{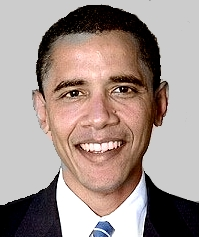}};
	\node at (4,2) (SG) {\includegraphics[width = 2cm]{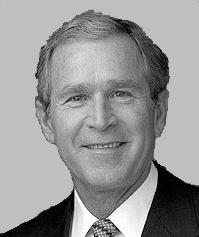}};
	\node at (4,-2) (SC) {\includegraphics[width = 2cm]{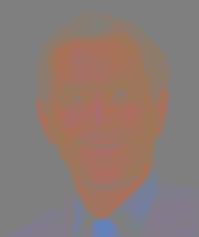}};
	\node[draw,line width=1mm,color = red,inner sep = 0] at (8,2) (TG) {\includegraphics[width = 2cm]{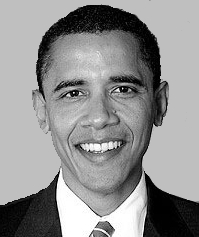}};
	\node at (8,-2) (TC) {\includegraphics[width = 2cm]{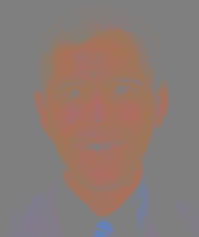}};
	\draw[<-] (SG.center) ..controls +(30:1)  and +(150:1).. (TG.center) node [pos = 0.5,label = 90:{Compute $\Phi$}](phicalc) {};
	\draw[<-] (SC.center) ..controls +(-30:1)  and +(-150:1).. (TC.center) node [pos = 0.5,label = -90:{Use $\Phi$}](phiuse) {};
	
	\node[text width = 2cm,align=center] at (2.1,0) (YUV1) {YUV\\decomposition};
	\node[text width = 2cm,align=center] at (9.9,0) (YUV2) {YUV\\inversion};
	\draw [myarrows,shorten <= 15](S)--(SG);
	\draw [myarrows,shorten <= 15](S)--(SC);
	\draw [myarrows,shorten <= 12](TG)--(R);
	\draw [myarrows,shorten <= 12](TC)--(R);
	
	\node[anchor = north,align = center] at (S.south) {Source\\image};
	\node[anchor = north,align = center] at (SG.south) {Luminance};
	\node[anchor = north,align = center] at (SC.south) {Source UV\\Chrominance};
	\node[anchor = north,align = center] at (TG.south) {Target\\image};
	\node[anchor = north,align = center] at (TC.south) {Mapped UV\\Chrominance};
	\node[anchor = north,align = center,outer sep = -1.5mm] at (R.south) {Colorized\\image};
	\end{tikzpicture}
	\end{TikZOrPDF}
	\caption[]{Overview of the color transfer. The mapping $\Phi$ is computed from Model~\eqref{main} 
	between the luminance channel of the source image and the target one. From this map, the chrominances of the source image are mapped. Finally, 
	from these chrominances and the target image the colorization result is computed.\footnotemark[1]}
	\label{overview}
\end{figure}

\paragraph{Chrominance Transfer by the Morphing Maps} 
Next we compute the morphing map between the two gray-scale images  
$I_{\text{temp}}$ and $I_{\text{tar}} $ with model~\eqref{main}. 
This results in the deformation sequence $\boldsymbol{\varphi}$ which produces the resulting map from the template image to the target one by concatenation
\begin{equation}\label{composition}
\Phi = \varphi_1 \circ \varphi_2 \circ ... \circ \varphi_{K}.
\end{equation}
Due to the discretization of the images, the map $\Phi$ is defined, for images of size $n_1\times n_2$, on the discrete grid $\mathcal{G} \coloneqq \{1,\dots,n_1\}\times \{1,\dots,n_2\}$:
\begin{equation}
\Phi :\mathcal{G} \rightarrow [1, n_1] \times  [1, n_2] , \quad
x \mapsto \Phi(x),
\end{equation}
where $\Phi(x)$ is the position  in the source image which corresponds to the pixel $x \in \mathcal{G}$ in the target image. 
Now we colorize the target image by computing its chrominance channels, denoted by $(U_{\text{tar}}(x),V_{\text{tar}}(x))$ at position $x$
as
\begin{equation} \label{uv_morph}
\big(U_{\text{tar}}(x),V_{\text{tar}}(x) \big) \coloneqq \big( U( \Phi(x) ),V(\Phi(x)) \big). 
\end{equation}
The chrominance channels of the target image are defined on the image grid $\mathcal{G}$, but usually $\Phi(x) \not \in \mathcal{G}$.
Therefore the values of the chrominance channels at $\Phi(x)$ have to be computed by interpolation. In our algorithm we use just bilinear interpolation
which is defined for $\Phi(x) = (p,q)$ with $(p,q) \in [i,i+1] \times [j,j+1]$, $(i,j) \in \{1,\ldots,n_1-1\} \times \{1,\ldots,n_2-1\}$ by
\begin{align} \label{bilinear}
U(\Phi(x)) = U (p,q) &\coloneqq (i+1-p, p-i) 
\begin{pmatrix} U(i,j)&U(i,j+1)\\ U(i+1,j)& U(i+1,j+1) \end{pmatrix}
\begin{pmatrix} j+1-q \\ q-j \end{pmatrix}.
\end{align}
Finally, we compute a colorized RGB image from its luminance 
$I_{\text{tar}}= Y_{\text{tar}}$ and the chrominance channels \eqref{uv_morph}
by the inverse of~\eqref{yuvrgb}: 
\begin{equation}\label{map_the_colors}
\left(\begin{array}{c}
R(x) \\ 
G(x) \\ 
B(x)
\end{array} \right) 
=
\left( \begin{array}{ccc}
1 & 0 & 1,13983 \\ 1 & -0,39465 & -0,58060 \\ 1 & 2,03211 & 0
\end{array} \right) 
\left(\begin{array}{c}
Y_{\text{tar}}(x) \\ 
U_{\text{tar}}(x)  \\ 
V_{\text{tar}}(x)
\end{array} \right) 
\end{equation}
for all $x \in \mathcal{G}$.\renewcommand*{\thefootnote}{\alph{footnote}}
Figure~\ref{overview}\footnotemark[1]\footnotetext[1]{Image segements with unified background of George W. Bush \url{https://commons.wikimedia.org/wiki/File:George-W-Bush.jpeg} and Barack Obama \url{https://commons.wikimedia.org/wiki/File:Barack_Obama.jpg}.} summarizes  our color transfer method.

\subsection{Variational Methods for Chrominance Postprocessing} \label{subsec32}

Sometimes the color transfer computed from the morphing map can be disturbed by artifacts. 
To improve the results, post-processing steps are usually applied in the literature.

Variational approaches are frequently applied in image colorization either directly or as a post-processing step,
see, e.g., \cite{levin2004colorization,pierre2015luminance,peter2016turning}. 
For instance, the technique of Gupta et al.~\cite{gupta2012image}  
uses the chrominance diffusion approach of Levin et al.~\cite{levin2004colorization}. 

In this paper, we build up on the model \cite{pierre2015luminance} 
suggested by one of the authors of that paper.
This variational model uses a functional with a specific regularization term to avoid ,,halo effects''. 
More precisely, the authors considered the minimizer of 
\begin{equation}\label{model_base}
\hat u = (\hat U, \hat V ) = \argmin_{(U,V)} \TV_{Y_{\text{tar}}} (U,V) + \alpha \int_{\Omega} | U(x) -  U_{\text{tar}}(x) |^2  + \,| V(x) -  V_{\text{tar}}(x) |^2 dx, 
\end{equation}
with 
\begin{equation}
\TV_{Y_{\text{tar}}}(U,V) \coloneqq \int_{\Omega} \sqrt{  \gamma | \nabla Y_{\text{tar}} |^2 
	+ | \nabla U |^2  + | \nabla V |^2 } \, dx.
\end{equation}
The first term in \eqref{model_base} is a coupled total variation term which enforces the chrominance channels to have a contour at the same location as the target gray-value image. 
The data fidelity term is the classical squared $L_2$-norm of the differences of the given and the desired chrominance channels. 
Note that the model in~\cite{pierre2015luminance} contains an additional box constraint. 
We apply the primal-dual Algorithm~\ref{algo_standard} to find the minimizer of the strictly convex  model~\eqref{model_base}. It uses an update on the step time parameters $\tau$ and $\sigma$, as proposed by Chambolle and Pock~\cite{chambolle2011first}, as well as a relaxation parameter $\theta$ to speed-up the convergence. 
Here we use the abbreviation $b \coloneqq (U_{\rm tar}, V_{\rm tar})$ and $u \coloneqq (U,V)$.
Further, $p$ is the dual variable which is pixel-wise in $\R^6$. 
The parameters  $\tau$ and $\sigma$ are intern time step sizes. 
The operator $\divv$ stands for the discrete divergence and $\nabla$ for the discrete gradient.
Further, the proximal mapping $P_{\mathcal{B}}$  is given pixel-wise, for $p\in \R^6$ by 
\begin{equation}\label{hatp}
P_{\mathcal{B}} \left(p\right) = \frac{\hat p }{\max \left(1 , \left\Vert \hat p  \right\Vert_2^2  \right)},
\;\;\;\text{ where } \;\;\;
\hat p \coloneqq p - \sigma \left( 
\begin{array}{c}
0 \\ 0 \\ 0 \\ 0 \\ \partial_x Y \\ \partial_y Y 
\end{array} \right).
\end{equation} 

\begin{algorithm}[htb]
	\caption{Minimization of~\eqref{model_base}.}\label{algo_standard}
	\begin{algorithmic}[1]
		\State $u^{0} \gets b $, $\overline{u}^{0} \gets  u^{0}$\vspace{0.15cm}
		\State $p^{0}\gets \nabla u^{0} $\vspace{0.15cm}
		\State $\sigma \gets 0.001$, \;\;$\tau \gets 20$\vspace{0.15cm}
		\For{ $n > 0$ }\vspace{0.15cm}
		\State $p^{n+1}\gets  P_{\mathcal{B}} \left( p^n + \sigma \nabla \overline{u}^{n} \right)$\vspace{0.15cm}
		\State $u^{n+1} \gets \dfrac{u^n + \tau \left( \divv(p^{n+1}) + \alpha b \right) }{1 + \tau \alpha} $ \vspace{0.15cm}
		\State $\theta =1 / \sqrt{1+\tau\alpha}$\vspace{0.15cm}
    \State $\tau = \theta\tau$ \;\; $\sigma = \sigma / \theta$\vspace{0.15cm}
		\State $\overline{u}^{n+1} \gets  u^{n+1} + \theta (u^{n+1} - u^n) $\vspace{0.15cm}
		\EndFor
		\vspace{0.15cm} \State $ \hat u \gets  u^{+\infty}. $\vspace{0.15cm}
	\end{algorithmic}
\end{algorithm} 

\begin{algorithm}[ht]
	\caption{Debiasing of Algorithm~\ref{algo_standard}.}\label{algo_unbias}
	\begin{algorithmic}[1]
		\State $u^0=b$, $\overline{u}^0=b$
		\vspace{0.15cm}\State $\delta \gets b - \hat u$
		\vspace{0.15cm}\State $\tilde u^0=\delta$, $\overline{\tilde u}^0=\delta$
		\vspace{0.15cm}\State $p^0\gets \nabla u$, $\tilde p^0\gets \nabla \tilde u$
		\vspace{0.15cm}\State $\sigma \gets 0.001$, \;\;$\tau \gets 20$
		\vspace{0.15cm}\For{ $n \geq 0$ }
		\vspace{0.15cm}\State $p^{n+1}\gets  P_{\mathcal{B}} \left( p^n + \sigma \nabla \overline{u}^{n} \right)$
		\vspace{0.15cm}\State $\tilde p^{n+1}\gets  \Pi_{p^n + \sigma \nabla \overline{u}^{n} } \left( \tilde p^n + \sigma \nabla \overline{\tilde u}^{n} \right)$
		\vspace{0.2cm}\State $u^{n+1} \gets  \dfrac{u^n + \tau \left( \divv(p^{n+1}) + \alpha b \right) }{1 + \tau \alpha}  $ 
		\vspace{0.2cm}\State $\tilde u^{n+1}\gets  \dfrac{\tilde u^n + \tau \left( \divv(\tilde p^{n+1}) + \alpha \delta \right) }{1 + \tau \alpha}  $ 
		\vspace{0.15cm}\State $\theta =1 / \sqrt{1+\tau\alpha}$
        \vspace{0.15cm}\State $\tau = \theta\tau$ \;\; $\sigma = \sigma / \theta$
		\vspace{0.15cm}\State $\overline{u}^{n+1} \gets  u^{n+1} + \theta (u^{n+1} - u^n)$
		\vspace{0.15cm}\State $\overline{\tilde u}^{n+1} \gets  \tilde u^{n+1} + \theta (\tilde u^{n+1} - \tilde u^n)$
		\vspace{0.15cm}\EndFor
		\vspace{0.2cm}\State $ \rho \gets \left\lbrace 
		\begin{matrix}
		\dfrac{\left\langle   \tilde u^{+\infty} | \delta \right\rangle}{\| \tilde u^{+\infty}\|_2^2 } & \text{if }  \tilde u^{+\infty} \not = 0\\
		1 & \text{otherwise.}
		\end{matrix} \right., $
		\vspace{0.15cm}\State $u_{\text{debiased}}  \gets \hat u + \rho \tilde u^{+\infty}. $
	\end{algorithmic}
\end{algorithm}

As mentioned in the work of Deledalle et al.~\cite{deledalle2017clear}, the minimization of the TV-$L_2$ model produces a biased result. 
This bias causes a lost of contrast in the case of gray-scale images, 
whereas it is visible as a lost of colorfulness in the case of model~\eqref{model_base}. 
The authors of~\cite{deledalle2017clear} describe an algorithm to remove such bias.
In this paper, we propose to modify this method for our model~\eqref{model_base} in order to enhance the result of Algorithm~\ref{algo_standard}. 
The final algorithm is summarized in Algorithm~\ref{algo_unbias}. Note that it uses the result $\hat u$ of Algorithm 1 as an input.
The proximal mapping $\Pi_p(\tilde p)$
within the algorithm is defined pixel-wise, for variables $p \in \R^6$ and   $\tilde p \in \R^6$, as

\begin{equation}\label{projection_pi}
\Pi_{p}\left( \tilde p \right) = \left\lbrace 
\begin{array}{ll}
\hat{\tilde p } &\text{ if } \| \hat p  \| < 1  
\\
\dfrac{1}{\| \hat p  \|} 
\left(
\hat{\tilde p } 
- 
\dfrac{
	\left\langle 
	\hat p ,\hat{\tilde p } 
	\right\rangle
}
{
	\| \hat p  \|^2 
} 
\hat p 
\right)
&\text{ otherwise,}
\end{array}
\right.
\end{equation}
where $\hat p$ and $\hat{\tilde p }  $ are defined as in~\eqref{hatp}.


\graphicspath{{./Images/}}
\begin{figure}[t]
	\centering
	\newcommand{\szw}{0.195\textwidth}
	\begin{subfigure}[t]{0.194\textwidth}
		\centering
		\includegraphics[width=0.98\textwidth]{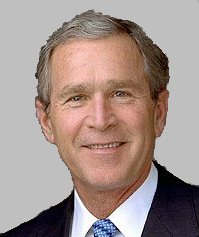}
		\caption{Source}
		\label{src}
	\end{subfigure}
	\begin{subfigure}[t]{0.194\textwidth}
		\centering
		\includegraphics[width=0.98\textwidth]{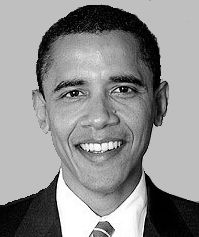}
		\caption{Target}
		\label{trg}
	\end{subfigure}
	\begin{subfigure}[t]{0.194\textwidth}
		\centering
		\includegraphics[width=0.98\textwidth]{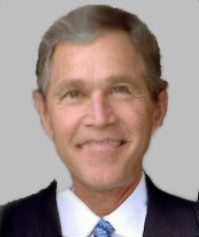}
		\caption{RGB morphing}
		\label{rgb}
	\end{subfigure}
	\begin{subfigure}[t]{0.194\textwidth}
		\centering
		\includegraphics[width=0.98\textwidth]{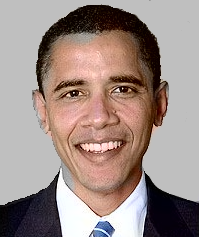}
		\caption{(UV) mapping}
		\label{uvm}
	\end{subfigure}
	\begin{subfigure}[t]{0.194\textwidth}
		\centering
		\includegraphics[width=0.98\textwidth]{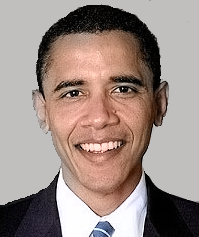}
		\caption{Final result}
		\label{rsl}
	\end{subfigure}
	\caption{
		Illustration of the colorization steps of our algorithm.
		\subref{rgb} The transport of the $R,G,B$ channel via the morphing map is not suited for colorization. 
		\subref{uvm} The result with our morphing method is already very good.
		\subref{rsl} It can be further improved by our variational post-processing.
		}
	\label{fig_steps}
\end{figure}

\graphicspath{{./Images/}}
\begin{figure}[tb]
	\centering
	\newcommand{\szw}{0.158\textwidth}
	\centerline{
		\begin{tabular}{c@{ }c@{ }c@{ }c@{ }c@{ }c}
			\includegraphics[width=\szw]{BuschObama_source} &
			\includegraphics[width=\szw]{BuschObama_target} &
			\includegraphics[width=\szw]{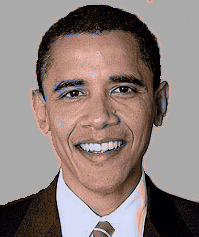} &
			\includegraphics[width=\szw]{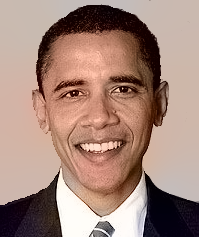} &
			\includegraphics[width=\szw]{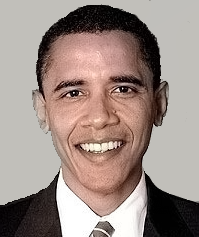} &
			\includegraphics[width=\szw]{BuschObama_our} \\
			\includegraphics[width=\szw]{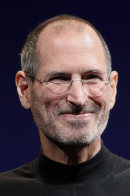} &
			\includegraphics[width=\szw]{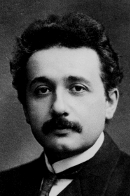} &
			\includegraphics[width=\szw]{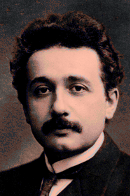} &
			\includegraphics[width=\szw]{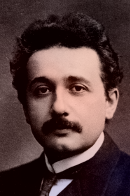} &
			\includegraphics[width=\szw]{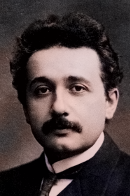} &
			\includegraphics[width=\szw]{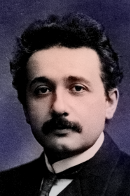} \\
			\includegraphics[width=\szw]{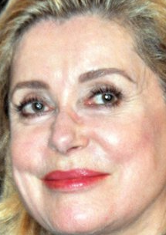} &
			\includegraphics[width=\szw]{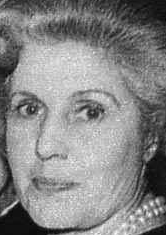} &
			\includegraphics[width=\szw]{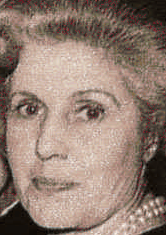} &
			\includegraphics[width=\szw]{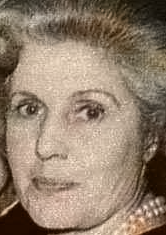} &
			\includegraphics[width=\szw]{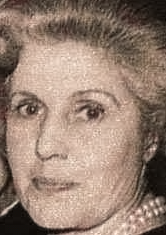} &
			\includegraphics[width=\szw]{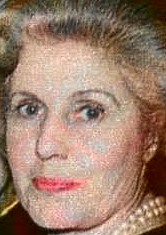} \\
			Source.&
			Target. &
				Welsh et al.~\cite{welsh2002transferring} &
				Gupta et al.~\cite{gupta2012image} &
				Pierre et al.~\cite{pierre2015luminance} &
			Our.
		\end{tabular}}
		\caption[]{Comparison of our approach with state-of-the-art methods on photographies. 
		 In contrast to these methods our model is not based on texture comparisons, 
		 but on the morphing of the shapes. Therefore it is able to handle faces images,
		where the background has frequently a similar texture as the skin.\footnotemark[2]}
		\label{numerical1}
	\end{figure}

	The results obtained at the different steps of the work-flow are presented for a particular image 
	in Figure~\ref{fig_steps}. 
	First we demonstrate in Figure~\ref{rgb} that simply transforming the $R,G,B$ channels via the morphing map 
	gives no meaningful results.
	Letting our morphing map act to the chrominance channels of our source image and applying \eqref{map_the_colors}
	with the luminance of our target image we get
	, e.g., Figure~\ref{uvm}) which is already a good result.
	However,  the forehead of Obama contains an artifact; a gray unsuitable color is visible here. 
		After a post-processing of the chrominance channels by our variational method the artifacts disappear as can be seen in Figure~\ref{rsl}.

	\section{Numerical Examples}\label{sec:numerical}
	
	\begin{table}
		\centering
		\begin{tabular}{lrrrr}
			\toprule
			Image & $\mu$ & K & $\gamma$ & $\alpha$ \\ 
			\midrule
			Fig.~\ref{numerical1}- 1. row& 0.025 & 24 & 50 & 0.005\\
			Fig.~\ref{numerical1}- 2. row& 0.05 & 24 & 25 & 0.005\\
			Fig.~\ref{numerical1}- 3. row& 0.05 & 12& - & -\\
			Fig.~\ref{numerical2}- 1. row& 0.005 & 32& -&-\\
			Fig.~\ref{numerical2}- 2. row& 0.0075 & 18& 50 & 0.05\\
			Fig.~\ref{numerical2}- 3. row& 0.04 & 24& -&-\\
			\bottomrule			
		\end{tabular}
		\caption{Parameters for our numerical experiments.\label{tab:params}}
	\end{table}
		
In this section, we compare our method with 
\begin{itemize}
\item[-]
the patch-based algorithm of Welsh et al.~\cite{welsh2002transferring}, 
\item[-] the patch-based method of Pierre et al.~\cite{pierre2015luminance}, and 
\item[-] the segmentation approach of Gupta et al.~\cite{gupta2012image}. 
\end{itemize}
We implemented our morphing algorithm in Matlab 2016b and used the Matlab intern function for the bilinear interpolation. 
The parameters are summarized in Table~\ref{tab:params}. Here $\lambda = \mu$ and $K$ are the parameters for the morphing step,
while $\gamma$ and $\alpha$ appear in the variational model for post-processing.
	
	First we compare the portraits in Figure \ref{numerical1}\footnotemark[2]\footnotetext[2]{Image segments of Steve Jobs \url{https://commons.wikimedia.org/wiki/File:Steve_Jobs_Headshot_2010-CROP.jpg}, Albert Einstein \url{https://commons.wikimedia.org/wiki/File:Einstein_Portr_05936.jpg}, Catherine Deneuve \url{https://commons.wikimedia.org/wiki/File:Catherine_Deneuve_2010.jpg} and Renée Deneuve \url{https://commons.wikimedia.org/wiki/File:Ren\%C3\%A9e_Deneuve.jpg}.}
	starting with the modern photographies in the first row.
	The approach of Welsh et al.~\cite{welsh2002transferring} is  based on a patch matching between images. 
	The patch comparison is done with basic texture descriptors (intensity of the central pixel and standard-deviation of the patches). 
	Since the background, as well as the skin are smooth, the distinction between them is unreliable if their intensities are similar. 
	Moreover, since no regularization is used after the color transfer, some artifacts occur. 
	For instance, some blue points appear on Obama's face, see Figure~\ref{numerical1}, first row. 
	The approach of Pierre et al.~\cite{pierre2015luminance} 
	is based on more sophisticated texture features for patches and applies a a variational model with total variation like regularization. 
		With this approach the artifacts mentioned above are less visible. Nevertheless, the forehead of Obama is purple which is unreliable. 
		The method of Gupta et al.~\cite{gupta2012image} uses texture descriptors after an over-segmentation, see, e.g., SLIC~\cite{achanta2012slic}. 
		The texture descriptors are based on SURF and Gabor features. 
		In the case of the Obama image, the descriptors are not able to distinguish the skin and other smooth parts, 
		leading to a background color different from the source image.
		
	The second and the third rows of Figure~\ref{numerical1} focus on the colorization of old photographies. 
	This challenging problem is a real application of image colorization which is sought, e.g.,  by the cinema industry. 
	Note that the texture of old images are disturbed by the natural grain which is not the case in modern photography.
	Thus, the texture comparison is unreliable for this application. 
	This issue is visible in all the comparison methods.
	For the portrait of Einstein the background is not colorized with the color of the source.
	Moreover, the color of the skin is different from those of the person in the source image. 
	For the picture of Deneuve, the color of her lips is not transferred to the target image (Deneuve's  mother) 
	with the state-of-the-art texture-based     algorithms. 
	With our morphing approach, the shapes of the faces are mapped. 
	Thus, the lips, as well as the eyes and the skin are well colorized with a color similar to the source image. 
	\\[1ex]	
	
\begin{figure}[t]
		\centering
		\newcommand{\szw}{0.158\textwidth}
			\begin{tabular}{c@{ }c@{ }c@{ }c@{ }c@{ }c}
				\includegraphics[width=\szw]{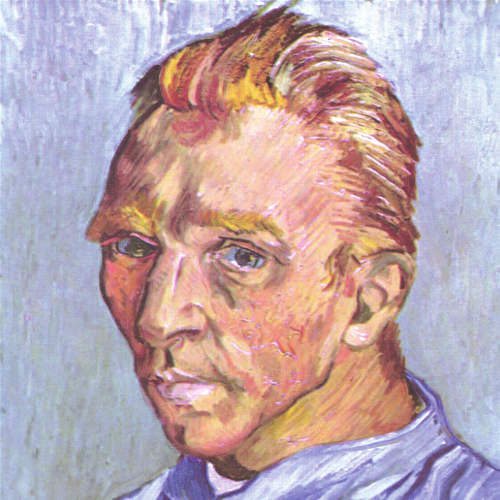} &
				\includegraphics[width=\szw]{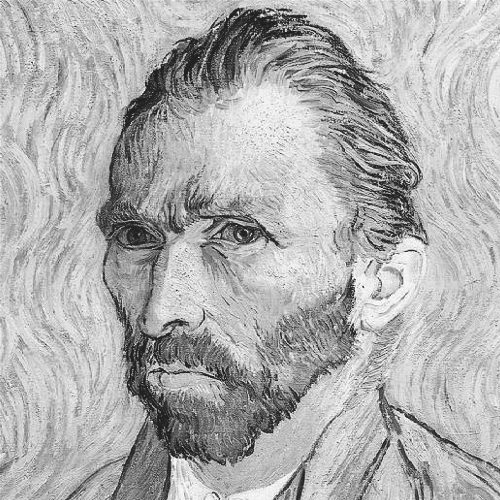} &
				\includegraphics[width=\szw]{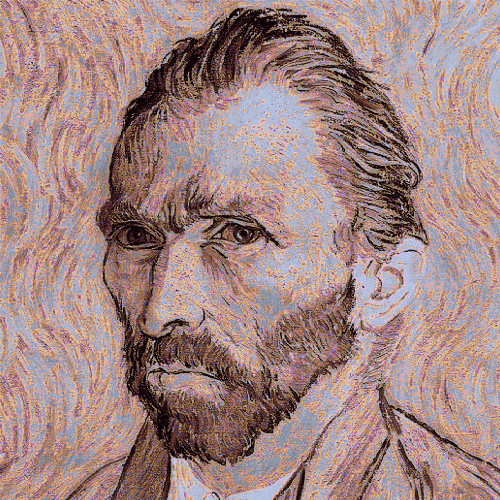} &
				\includegraphics[width=\szw]{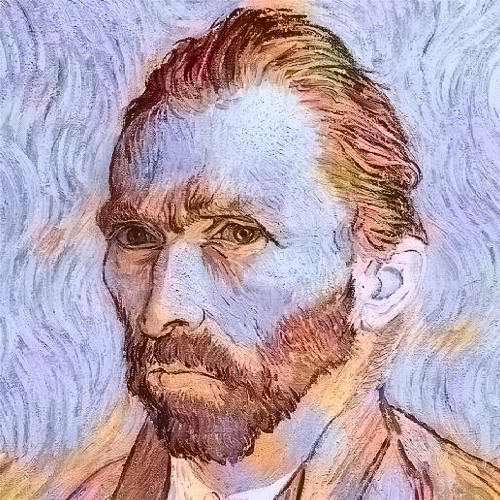} &
				\includegraphics[width=\szw]{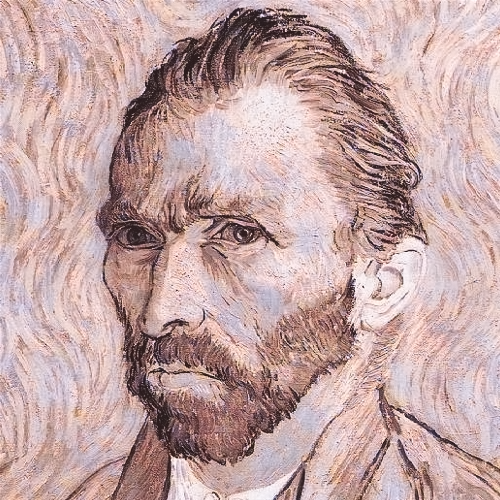} &
				\includegraphics[width=\szw]{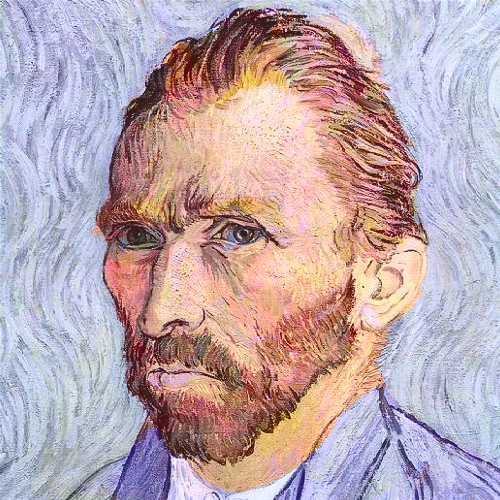} \\
				\includegraphics[width=\szw]{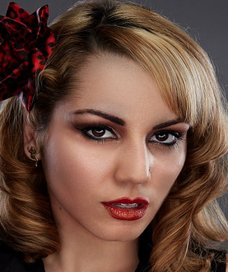} &
				\includegraphics[width=\szw]{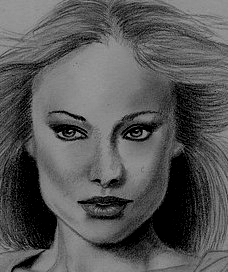} &
				\includegraphics[width=\szw]{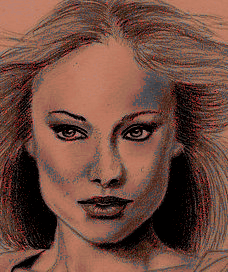} &
				\includegraphics[width=\szw]{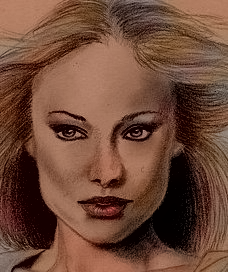} &
				\includegraphics[width=\szw]{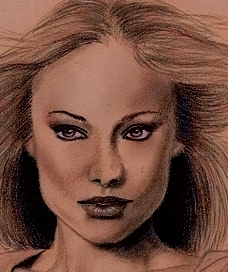} &
				\includegraphics[width=\szw]{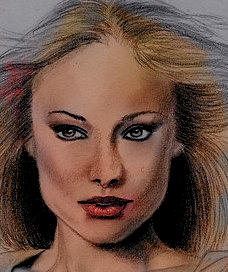} \\
				\includegraphics[width=\szw]{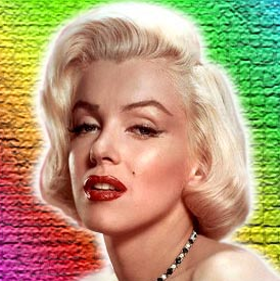} &
				\includegraphics[width=\szw]{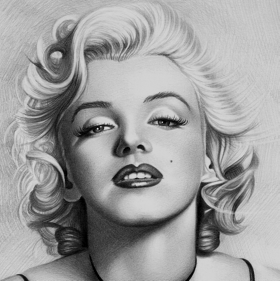} &
				\includegraphics[width=\szw]{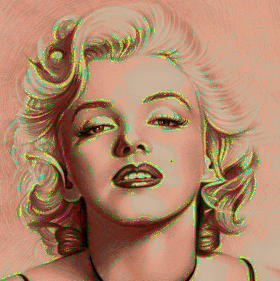} &
				\includegraphics[width=\szw]{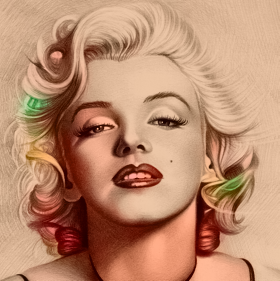} &
				\includegraphics[width=\szw]{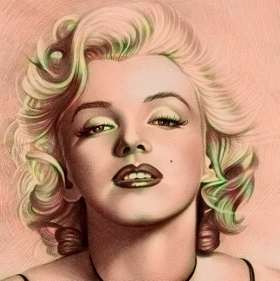} &
				\includegraphics[width=\szw]{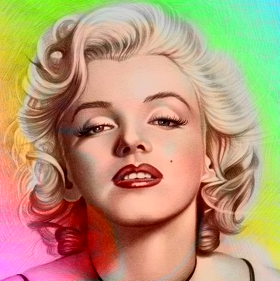} \\
				Source.&
				Target. &
				Welsh et al.~\cite{welsh2002transferring} &
				Gupta et al.~\cite{gupta2012image} &
				Pierre et al.~\cite{pierre2015luminance} &
				Our.
			\end{tabular}
			\caption[]{Results including painted faces. Only our morphing method is able to colorize the target images
			in an authentic way.\footnotemark[3]}
			\label{numerical2}
		\end{figure}

\begin{figure}[t]
			\centering
			\begin{tabular}{c@{ }c@{ }c@{ }c@{ }c}
				\includegraphics[width=.19\textwidth]{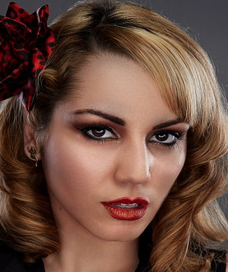} &
				\includegraphics[width=.19\textwidth]{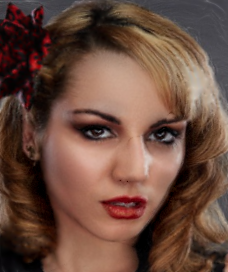} &
				\includegraphics[width=.19\textwidth]{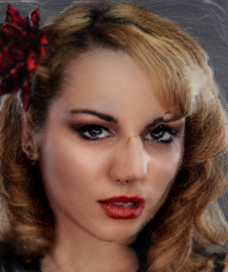} &
				\includegraphics[width=.19\textwidth]{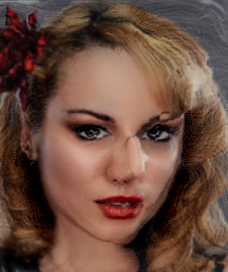} &			
				\includegraphics[width=.19\textwidth]{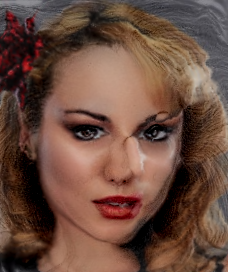} \\				
				\includegraphics[width=.19\textwidth]{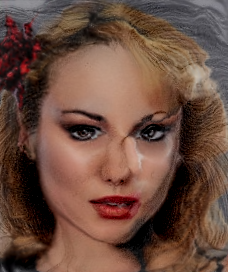} &
				\includegraphics[width=.19\textwidth]{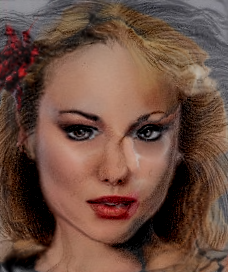} &
				\includegraphics[width=.19\textwidth]{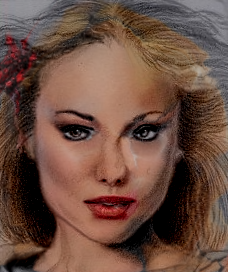} &
				\includegraphics[width=.19\textwidth]{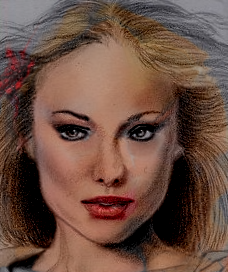} &			
				\includegraphics[width=.19\textwidth]{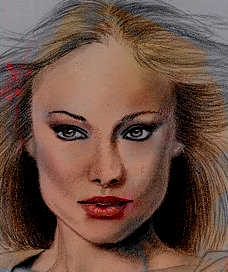} 
			\end{tabular}
			\caption{Color transport along the image path.\label{fig:imagepath}}
		\end{figure}

	In Figure~\ref{numerical2}\footnotemark[3]\footnotetext[3]{Image segments of self-portraits of Vincent van Gogh \url{https://commons.wikimedia.org/wiki/File:Vincent_Willem_van_Gogh_102.jpg} and \url{https://commons.wikimedia.org/wiki/File:Vincent_van_Gogh_-_Self-Portrait_-_Google_Art_Project.jpg},\\ a photography of a woman \url{https://pixabay.com/en/woman-portrait-face-model-canon-659352/}, a drawing of a woman \url{https://pixabay.com/en/black-and-white-drawing-woman-actor-1724363/},\\ a color image of Marilyn Monroe \url{https://www.flickr.com/photos/7477245@N05/5272564106} created by Luiz Fernando Reis, and a drawing of Marilyn Monroe \url{https://pixabay.com/en/marilyn-monroe-art-draw-marilyn-885229/}}, we provide results including painted images. Note that we use the same Van Gogh self-portraits as in \cite{BER15}. Due to the low contrast of the ear and suit to the background we add here the same segmentation information as in \cite{BER15}, which means our images are two dimensional during the calculation for the results shown in the first row of Figure~\ref{numerical2}.
	In these examples the similarity of the shapes between the source and target images is again more reliable than the matching of the textures so that only our morphing approach produces suitable results. Consider in particular the lips of the woman in the second and third row. A non post-processed result for the woman in the second row is shown in Figure~\ref{fig:imagepath}. Comparing the two images we see that only small details change but most of the colorization is done by the morphing.

As morphing approach calculates at the same time an image and diffeomorphism path, 
we can not only show the final colorization result, but also the way the color is transported along this path. 
We illustrate this by the second row of Figure~\ref{numerical2}, where every second image along the path is shown.
In Figure~\ref{fig:imagepath} we see how the UV channels are transported via the diffeomorphism path of
the Y channels of the source and the target images.  We see that even though the right part of the images undergoes large transformations, 
the eyes and the mouth are moved to the correct places. Note that the final image does not contain a post-processing in contrast to those in 
the second row of Figure~\ref{numerical2}. However, the result is already quite good.

		\section{Conclusion } \label{sec:conclusions}
		%
		In this paper, we propose an finite difference method to compute a morphing map between a source and a target gray-value image.
		This map enables us to transfer the color from the source image to the target one, based on the shapes of the faces. 
		An additional post-processing step, based on a variational model with a luminance guided  total variation regularization 
		and an update to make the result unbiased may be added to remove some possible artifacts. 
		The results are very convincing and outperform state-of-the-art approaches on historical photographies. 
		
		Let us notice some special issues of our approach in view of an integration into a more global framework for an exemplar-based image and video colorization. 
First of all, our method works well on faces and on object with similar shapes, but when this hypothesis is not fulfilled, some artifacts can appear. 
Therefore, instead of using our algorithm on full images, a face detection algorithm can be used to focus on the face colorization.
Let us remark that faces in image can be detected by efficient, recent methods, see, e.g.,~\cite{chen2016supervised}. 
In future work, the method will be integrated into a complete framework for exemplar-based image and video colorization. 

Second, the morphing map has to be computed between images with the same size. 
This issue can be easily solved with a simple interpolation of the images. 
Keeping the ratio between the width and the height of faces similar, 
the distortion produced by such interpolation is small enough to support our method. 

\appendix
\section{}\label{app}	

	Fixing $\zb v$, resp., $\boldsymbol{\varphi}$ leads to the image sequence problem \eqref{fix_u}.
	In the following we show how this can be solved via the linear system of equations
	arising from Euler-Lagrange equation of the functional.
	
The terms in $\mathcal{J}_{\boldsymbol{\varphi}}(\zb I)$ containing $I_k$ are
\begin{equation}
\int_\Omega \lvert I_k \left( \varphi_{k+1} (\zb x) \right) - I_{k+1}(\zb x) \rvert^2 
+ \lvert I_k(\zb x) - I_{k-1}\left( \varphi_k (\zb x) \right)\rvert^2 \, \dx \zb x.
\end{equation}
Noting that $\varphi_{k+1}$ is a diffeomorphism, 
the variable transform $\zb y \coloneqq \varphi_{k+1} (\zb x)$ gives
\begin{align}
&\int_\Omega \lvert I_k(\zb x) - I_{k+1}\left( \varphi_{k+1}^{-1} (\zb x) \right) \rvert^2 
\lvert \operatorname{det} \nabla \varphi_{k+1}^{-1} \left( \varphi_{k+1}^{-1}(\zb x) \right) \rvert^2
+ 
\lvert I_{k}(\zb x) - I_{k-1}\left( \varphi_{k} (\zb x) \right) \rvert^2 \, \dx \zb x\\
&= 
\int_\Omega \lvert I_k(\zb x) - I_{k+1}\left( \varphi_{k+1}^{-1} (\zb x) \right) \rvert^2 
\lvert \operatorname{det} \nabla \varphi_{k+1} \left( \varphi_{k+1}^{-1}(\zb x) \right) \rvert^{-1}
+ 
\lvert I_{k}(\zb x) - I_{k-1}\left( \varphi_{k} (\zb x) \right) \rvert^2 \, \dx \zb x
\end{align}
Setting
\begin{equation} \label{a}
a_k\left(\varphi_{k+1}^{-1}(\zb x) \right) \coloneqq \lvert \operatorname{det} \nabla \varphi_{k+1} \left( \varphi_{k+1}^{-1}(\zb x) \right) \rvert^{-1},
\end{equation}
the Euler-Lagrange equations read for $k=1,\ldots,K-1$ as
\begin{align}
\bigl( I_k(\zb x) - I_{k+1}(\varphi_{k+1}^{-1}(\zb x)) \bigr) 
a_k\left(\varphi_{k+1}^{-1}(\zb x) \right)
+ 
I_{k}(\zb x) - I_{k-1}\bigl(\varphi_{k}( \zb x)\bigr)  &= 0, \\
- I_{k-1}\bigl(\varphi_{k}( \zb x)\bigr) + \left(1+ a_k\left(\varphi_{k+1}^{-1}(\zb x) \right) \right) I_{k}(\zb x) 
- a_k \left(\varphi_{k+1}^{-1}(\zb x) \right) I_{k+1}(\varphi_{k+1}^{-1}(\zb x))&= 0.
\label{el}
\end{align}
We introduce
$$
X_K(\zb x)\coloneqq \zb x, \quad X_{k-1}(\zb x) \coloneqq \varphi_k \bigl(X_{k}(\zb x)\bigr), \quad k=1, \ldots,K,
$$
which can be computed for each $\zb x \in \Omega$ since the $\varphi_k$, $k=1,\ldots,K$ are given.
Then \eqref{el} can be rewritten for $k=1,\ldots,K-1$ as
\begin{equation} \label{system_1}
- I_{k-1}\left(X_{k-1}( \zb x)\right) + (1+ a_k\left( X_{k+1}(\zb x) \right) I_{k}\left( X_k(\zb x) \right) 
- a_k\left( X_{k+1}(\zb x) \right) I_{k+1}\left(X_{k+1}(\zb x)\right)= 0. 
\end{equation}
In matrix-vector form, this reads with 
$$
F_0 = I_{\textrm{temp}} \left( X_0(\zb x) \right), \quad 
F_K = I_{\textrm{ref}} (\zb x),\quad 
F_k \coloneqq  I_{k}\left( X_k(\zb x) \right), \quad k=1,\ldots, K-1
$$ 
for fixed $\zb x \in \Omega$
and 
$F \coloneqq (F_1,\ldots,F_{K-1})^\tT$ 
as
\begin{equation} \label{eq:tridiag}
A \, F  = (F_0, 0, \ldots, 0, a_{K-1} F_K)^\tT, \quad  A \coloneqq \text{tridiag} (-1, 1+a_k, -a_k)_{k=1}^{K-1}. 
\end{equation}
Assuming that $a_k >0$ which is the case in practical applications, 
the matrix $A$ is irreducible diagonal dominant and thus invertible.

\section*{Acknowledgments}
{Funding by the German Research Foundation (DFG) within the project STE 571/13-1 is gratefully acknowledged.}

\bibliographystyle{plain}
\bibliography{references}

\begin{thebibliography}{10}

\bibitem{achanta2012slic}
Radhakrishna Achanta, Appu Shaji, Kevin Smith, Aurelien Lucchi, Pascal Fua, and
  Sabine S{\"u}sstrunk.
\newblock Slic superpixels compared to state-of-the-art superpixel methods.
\newblock {\em IEEE Transactions on Pattern Analysis and Machine Intelligence},
  34(11):2274--2282, 2012.

\bibitem{TY2005}
Trouv\'e Alian and Laurent Younes.
\newblock Local geometry of deformable templates.
\newblock {\em SIAM Journal of Mathematical Analysis}, 37(2):17--59, 2005.

\bibitem{BER15}
Benjamin Berkels, Alexander Effland, and Martin Rumpf.
\newblock Time discrete geodesic paths in the space of images.
\newblock {\em SIAM Journal on Imaging Sciences}, 8(3):1457--1488, 2015.

\bibitem{chambolle2011first}
Antonin Chambolle and Thomas Pock.
\newblock A first-order primal-dual algorithm for convex problems with
  applications to imaging.
\newblock {\em Journal of Mathematical Imaging and Vision}, 40(1):120--145,
  2011.

\bibitem{charpiat2010automatic}
G.~Charpiat, M.~Hofmann, and B.~Sch\"{o}lkopf.
\newblock Automatic image colorization via multimodal predictions.
\newblock In {\em European Conference on Computer Vision}, pages 126--139.
  Springer, 2008.

\bibitem{chen2016supervised}
Dong Chen, Gang Hua, Fang Wen, and Jian Sun.
\newblock Supervised transformer network for efficient face detection.
\newblock In {\em European Conference on Computer Vision}, pages 122--138.
  Springer, 2016.

\bibitem{chen2004grayscale}
Tongbo Chen, Yan Wang, Volker Schillings, and Christoph Meinel.
\newblock Grayscale image matting and colorization.
\newblock In {\em Asian Conference on Computer Vision}, pages 1164--1169, 2004.

\bibitem{chia2011semantic}
Alex Yong-Sang Chia, Shaojie Zhuo, Raj~Gupta Kumar, Yu-Wing Tai, Siu-Yeung Cho,
  Ping Tan, and Stephen Lin.
\newblock Semantic colorization with internet images.
\newblock In {\em ACM SIGGRAPH ASIA}, 2011.

\bibitem{deledalle2017clear}
Charles-Alban Deledalle, Nicolas Papadakis, Joseph Salmon, and Samuel Vaiter.
\newblock Clear: Covariant least-square refitting with applications to image
  restoration.
\newblock {\em SIAM Journal on Imaging Sciences}, 10(1):243--284, 2017.

\bibitem{DGM98}
Paul Dupuis, Ulf Grenander, and Michael~I. Miller.
\newblock Variational problems on flows of diffeomorphisms for image matching.
\newblock {\em Quarterly of Applied Mathematics}, 56(3):587--600, 1998.

\bibitem{efros1999texture}
Alexei~A Efros and Thomas~K Leung.
\newblock Texture synthesis by non-parametric sampling.
\newblock In {\em IEEE International Conference on Computer Vision}, volume~2,
  pages 1033--1038, 1999.

\bibitem{gupta2012image}
Raj~Kumar Gupta, Alex Yong-Sang Chia, Deepu Rajan, Ee~Sin Ng, and Huang
  Zhiyong.
\newblock Image colorization using similar images.
\newblock In {\em ACM International Conference on Multimedia}, pages 369--378,
  2012.

\bibitem{HM06}
Eldad Haber and Jan Modersitzki.
\newblock A multilevel method for image registration.
\newblock {\em SIAM Journal on Scientific Computing}, 27(5):1594--1607, 2006.

\bibitem{hertzmann2001image}
Aaron Hertzmann, Charles~E Jacobs, Nuria Oliver, Brian Curless, and David~H
  Salesin.
\newblock Image analogies.
\newblock In {\em Proceedings of the 28th annual conference on Computer
  graphics and interactive techniques}, pages 327--340. ACM, 2001.

\bibitem{irony2005colorization}
Revital Irony, Daniel Cohen-Or, and Dani Lischinski.
\newblock Colorization by example.
\newblock In {\em Eurographics conference on Rendering Techniques}, pages
  201--210. Eurographics Association, 2005.

\bibitem{levin2004colorization}
Anat Levin, Dani Lischinski, and Yair Weiss.
\newblock Colorization using optimization.
\newblock In {\em ACM Transactions on Graphics}, volume 23--3, pages 689--694,
  2004.

\bibitem{MY2001}
Michael~I. Miller and Laurent Younes.
\newblock Group actions, homeomorphisms, and matching: A general framework.
\newblock {\em International Journal of Computer Vision}, 41(1-2):61--84, 2001.

\bibitem{Mod2004}
Jan Modersitzki.
\newblock {\em Numerical Methods for Image Registration}.
\newblock Oxford University Press on Demand, 2004.

\bibitem{PS2017}
Johannes Persch and Gabriele Steidl.
\newblock Morphing of manifold-valued images.
\newblock {\em in preparation}, 2017.

\bibitem{peter2016turning}
Pascal Peter, Lilli Kaufhold, and Joachim Weickert.
\newblock Turning diffusion-based image colorization into efficient color
  compression.
\newblock {\em IEEE Transactions on Image Processing}, 2016.

\bibitem{pierre2015luminance}
Fabien Pierre, Jean-Fran{\c c}ois Aujol, Aur{\'e}lie Bugeau, Nicolas Papadakis,
  and Vinh-Thong Ta.
\newblock Luminance-chrominance model for image colorization.
\newblock {\em SIAM Journal on Imaging Sciences}, 8(1):536--563, 2015.

\bibitem{welsh2002transferring}
Tomihisa Welsh, Michael Ashikhmin, and Klaus Mueller.
\newblock Transferring color to greyscale images.
\newblock In {\em ACM Transactions on Graphics}, volume 21--3, pages 277--280,
  2002.

\bibitem{Younes2010}
Laurent Younes.
\newblock {\em Shapes and Diffeomorphisms}.
\newblock Springer-Verlag, Berlin, 2010.

\bibitem{zhang2016colorful}
Richard Zhang, Phillip Isola, and Alexei~A Efros.
\newblock Colorful image colorization.
\newblock In {\em European Conference on Computer Vision}, pages 1--16.
  Springer, 2016.

\end{thebibliography}


\end{document}